\documentclass[sigconf]{acmart}
\usepackage{hyperref}
\usepackage{booktabs} 
\usepackage{enumerate}
\usepackage{subfigure}
\usepackage{bigstrut}
\usepackage{makecell}
\usepackage{multirow}

\copyrightyear{2019}
\acmYear{1997}
\setcopyright{acmcopyright}
\acmConference[DAC '19]{The 56th Annual Design Automation Conference 2019}{June 2--6, 2019}{Las Vegas, NV, USA}
\acmBooktitle{The 56th Annual Design Automation Conference 2019 (DAC '19), June 2--6, 2019, Las Vegas, NV, USA}
\acmPrice{15.00}
\acmDOI{10.1145/3316781.3317820}
\acmISBN{978-1-4503-6725-7/19/06}

\begin{document}
\title{eSLAM: An Energy-Efficient Accelerator for Real-Time ORB-SLAM on FPGA Platform}
\titlenote{This work was supported in part by the National Natural Science Foundation of China (61602022, 61501013, 61571023, 61521091 and 1157040329), State Key Laboratory of Software Development Environment (SKLSDE-2018ZX-07), National Key Technology Program of China (2017ZX01032101), CCF-Tencent IAGR20180101 and the International Collaboration Project under Grant B16001.}

\author{Runze Liu $^{\dag\P}$,\hspace{1em} Jianlei Yang $^{\dag\P}$,\hspace{1em} Yiran Chen $^\S$,\hspace{1em} Weisheng Zhao $^{\ddag\P}$}
\affiliation{
   \institution{$^\dag$ School of Computer Science and Engineering, Beihang University, Beijing, 100191, China.}
   \institution{$^\ddag$ School of Electronic and Information Engineering, Beihang University, Beijing, 100191, China.}
   \institution{$^\P$ Beijing Advanced Innovation Center for Big Data and Brain Computing, Beihang University, Beijing, 100191, China.}
   \institution{$^\S$ Department of Electrical and Computer Engineering, Duke University, Durham, NC 27708, USA.}
}
\affiliation{
    \institution{jianlei@buaa.edu.cn \hspace{2em} weisheng.zhao@buaa.edu.cn}
}

\begin{abstract}
Simultaneous Localization and Mapping (SLAM) is a critical task for autonomous navigation. However, due to the computational complexity of SLAM algorithms, it is very difficult to achieve real-time implementation on low-power platforms. We propose an energy-efficient architecture for real-time ORB (Oriented-FAST and Rotated-BRIEF) based visual SLAM system by accelerating the most time-consuming stages of feature extraction and matching on FPGA platform. Moreover, the original ORB descriptor pattern is reformed as a rotational symmetric manner which is much more hardware friendly. Optimizations including rescheduling and parallelizing are further utilized to improve the throughput and reduce the memory footprint. Compared with Intel i7 and ARM Cortex-A9 CPUs on TUM dataset, our FPGA realization achieves up to $3\times $ and $31\times $ frame rate improvement, as well as up to $71\times $ and $25\times $ energy efficiency improvement, respectively.

\end{abstract}


\maketitle

\section{INTRODUCTION}\label{section:introduction}
Simultaneous Localization and Mapping (SLAM)~\cite{durrant2006simultaneous} is a critical technique for autonomous navigation systems to build/update a map of the surrounding environment and estimate their own locations in this map. SLAM is a fundamental problem for higher-level tasks such as path planning and navigation, and widely used in applications such as self-driving cars, robotics, virtual reality and augmented reality.

Recently, feature-based visual SLAM has received particular attention because of its robustness to large motions and illumination changes compared with other visual SLAM approaches such as optical flow method or direct method. Among feature-based approaches, ORB (Oriented-FAST and Rotated-BRIEF)~\cite{Rublee2012ORB} is the most widely adopted feature because of its high efficiency and robustness. However, the high computational intensity of feature extraction and matching makes it very challenging to run ORB-based visual SLAM on low-power embedded platforms, such as drones and mobile robots, for real-time applications.

Several prior efforts have been made to accelerate visual SLAM on low-power platforms, but no fully integrated ORB-based visual SLAM is proposed on such platforms so far. Feature matching and ORB extraction is accelerated on FPGA for visual SLAM system, respectively in \cite{Cong2011Accelerating} and \cite{Fang2018FPGA}. A SIFT-feature based SLAM is implemented on FPGA~\cite{Gu2015An} where only matrix computation is accelerated but the most time-consuming part, feature extraction, is not involved. A optical-flow based visual inertial odometry is implemented on ASIC~\cite{suleiman2018navion}, which is relatively less computational intensive but may fail in scenarios with variational illuminations or large motions/displacements, because the basic assumptions of optical flow method are invalid in these scenarios~\cite{fleet2006optical}.

In this paper, \texttt{eSLAM} is proposed as a heterogeneous architecture of ORB-based visual SLAM system. The most time-consuming procedures of feature extraction and matching are accelerated on FPGA while the remaining tasks including pose estimation, pose optimization and map updating are performed on the host ARM processor. The main contributions of this paper are listed as below:
\begin{itemize}
\item A novel ORB-based visual SLAM accelerator is proposed for real-time applications on energy-efficient FPGA platforms.
\item A rotationally symmetric ORB descriptor pattern is utilized to make our algorithm much more hardware-friendly.
\item Optimization including rescheduling and parallelizing are further exploited to improve the computation throughput.
\end{itemize}

The remainder of this paper is organized as follows. Section \ref{section:ORB-SLAM} presents the ORB-based visual SLAM framework and the introduced rotationally symmetric descriptor. Section \ref{section:architecture} illustrates the detailed architecture of \texttt{eSLAM}. Experimental results are evaluated for the proposed \texttt{eSLAM} in Section \ref{section:evaluation_and_results}. Concluding remarks are given in Section \ref{section:conclusion}.

\section{ORB-BASED VISUAL SLAM SYSTEM}\label{section:ORB-SLAM}

\subsection{ORB-SLAM Framework}

\begin{figure}
    \centering
    \includegraphics[width=0.4\textwidth]{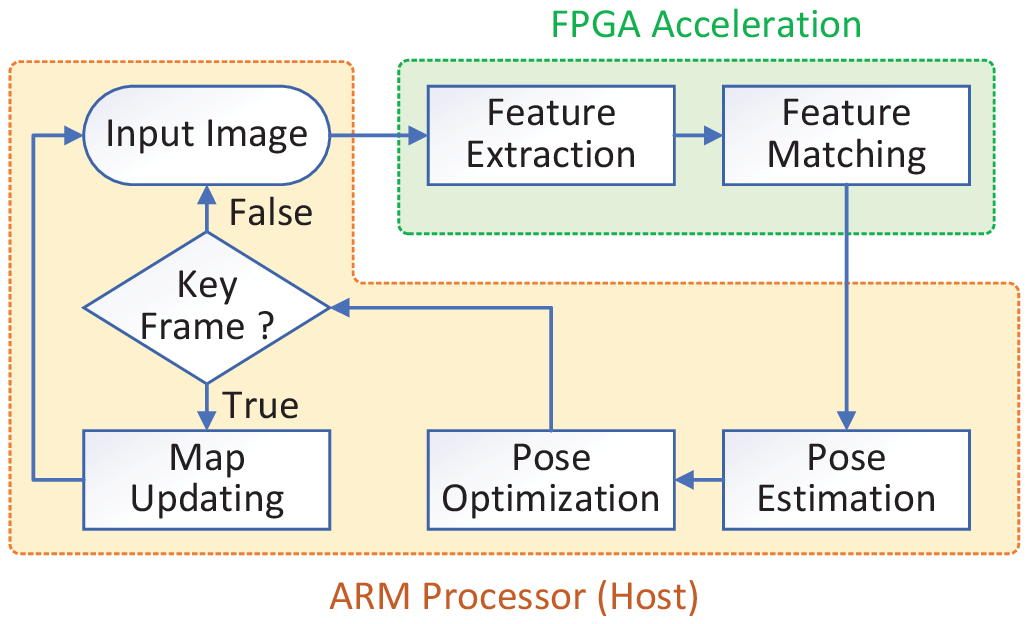}
    \caption{Visual SLAM algorithm framework.}
    \label{figure:overflow}
\end{figure}

The ORB-based visual SLAM system takes RGB-D (RGB and depth) images for mapping and localization. Its framework, as shown in Figure \ref{figure:overflow}, consists of five main procedures: feature extraction, feature matching, pose estimation, pose optimization and map updating. In this work, feature extraction and feature matching are accelerated on FPGA, and remaining tasks are performed on ARM processor.

\textbf{Feature Extraction:} In this function, ORB features are extracted from the input RGB images. ORB is a very efficient and robust combination of FAST (Features from Accelerated Segment Test) keypoint and BRIEF (Binary Robust Independent Elementary Features)~\cite{Calonder2010BRIEF} descriptor. It calculates orientations of every feature and rotates the descriptor pattern accordingly to make the features rotationally invariant. And to obtain scale invariance, a 4-layer pyramid is generated from the original image. Aiming to implement ORB algorithm on hardware efficiently, a hardware-friendly, rotationally symmetric BRIEF descriptor pattern is proposed in this work and illustrated in Section \ref{sec:rsbrief}.

\textbf{Feature Matching:} In feature matching, each feature detected in the current frame is matched with a 3D map point in the global map according to the distance between their BRIEF descriptors. BRIEF descriptors are binary strings, their distances are described by Hamming distances.

\textbf{Pose Estimation:} We apply PnP (Perspective-n-Points) method to the matched feature pairs to estimate the translation and the rotation of the camera. RANSAC (Random Sample Consensus) is used to eliminate the mismatches.

\textbf{Pose Optimization:} In this function, camera pose estimated by PnP is optimized by minimizing the reprojection error of the observed map points. Assuming that the pixel coordinates of the features in the current frame are $(c_{1}, c_{2}, ..., c_{n})$, the positions of the matched map points are $(g_{1}, g_{2}, ..., g_{n})$, the pose of the camera is $p$, and $h(g_{i}, p)$ refers to the pixel coordinate of $g_{i}$ when it is projected to the current frame. The reprojection error $E$ can be defined as the following formula:
\begin{equation}
E = \sum_{i = 1}^{n}\left \| c_i-h(g_i,p) \right \|^2
\end{equation}
Levenberg-Marquardt method~\cite{Mor1978The} is applied iteratively to minimize $E$ while adjusting the camera pose $p$.

\textbf{Map Updating:} Map updating is only executed in key frames. Key frames are a set of frames where the translation or rotation of the camera is larger than a threshold. When a key frame is detected, the 3D map points in the key frame are added to the global map, and the map points that have not been matched for a long period of time are deleted from the global map to prevent it from becoming too large.

\subsection{Rotationally Symmetric BRIEF}\label{sec:rsbrief}

To compute the BRIEF descriptor of a feature, 2 sets of locations in the neighborhood around the feature, $L_S(S_1,S_2,$ $...,S_{256})$ and $L_D (D_1,D_2,$ $...,D_{256})$, are introduced. In ORB algorithm, to make the feature invariant to rotation, $L_S$ and $L_D$ are rotated according to the feature's orientation and denoted as $L_{SR}(SR_1, SR_2,$ $..., SR_{256})$ and $L_{DR}(DR_1, DR_2, ...,$ $DR_{256})$. The descriptor $T(B_1,B_2,$ $...,B_{256})$ is a $256$ bits binary string. $B_i$ is $1$ if $I(SR_i) > I(DR_i)$, else it is $0$, where $I(SR_i)$ and $I(DR_i)$ are the pixel intensities on location $SR_i$ and $DR_i$.

\begin{figure}
    \centering
    \subfigure{
        \label{figure:rs_brief}
        \includegraphics[width=0.22\textwidth]{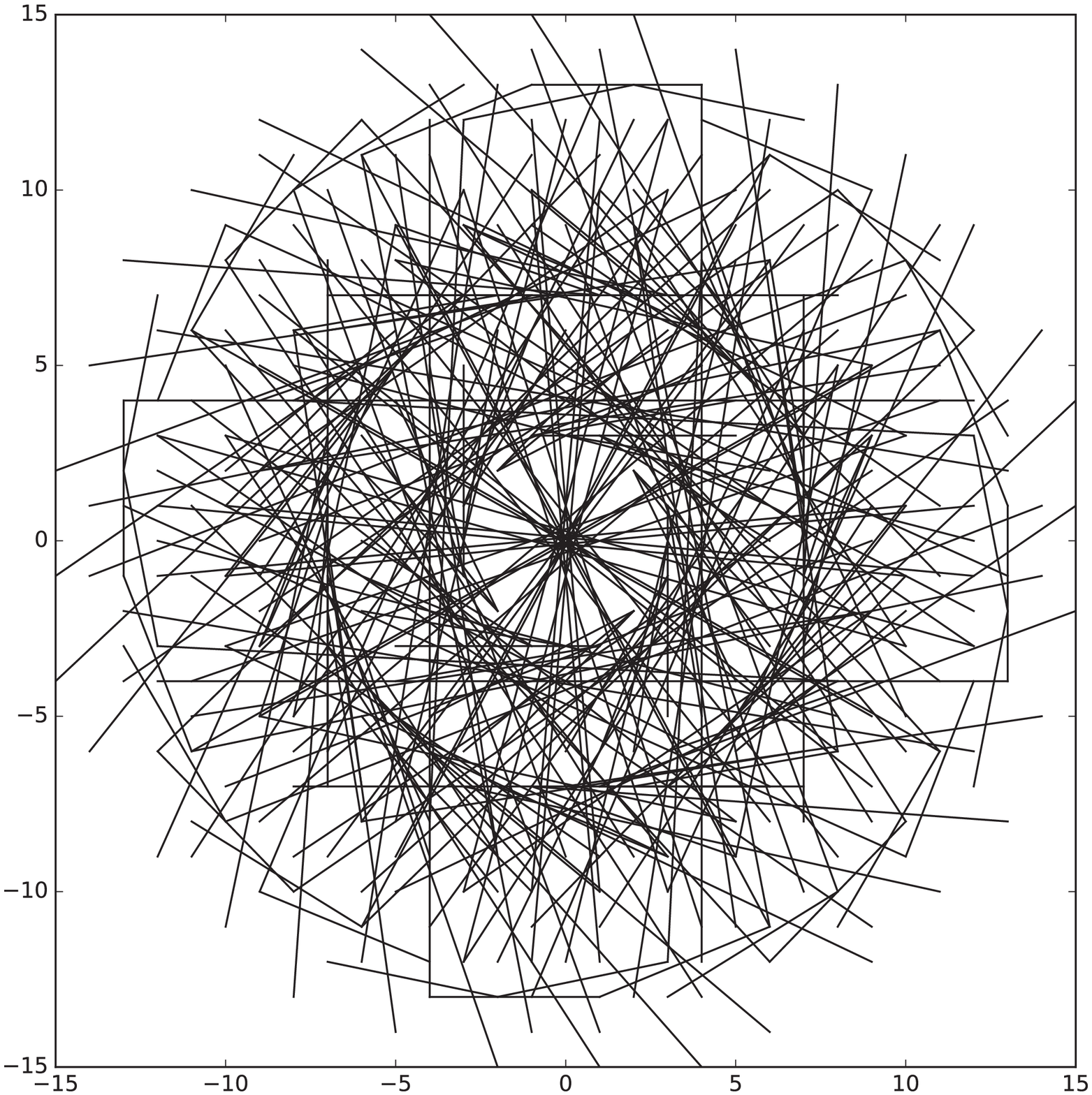}}
    \subfigure{
        \label{figure:brief}
        \includegraphics[width=0.22\textwidth]{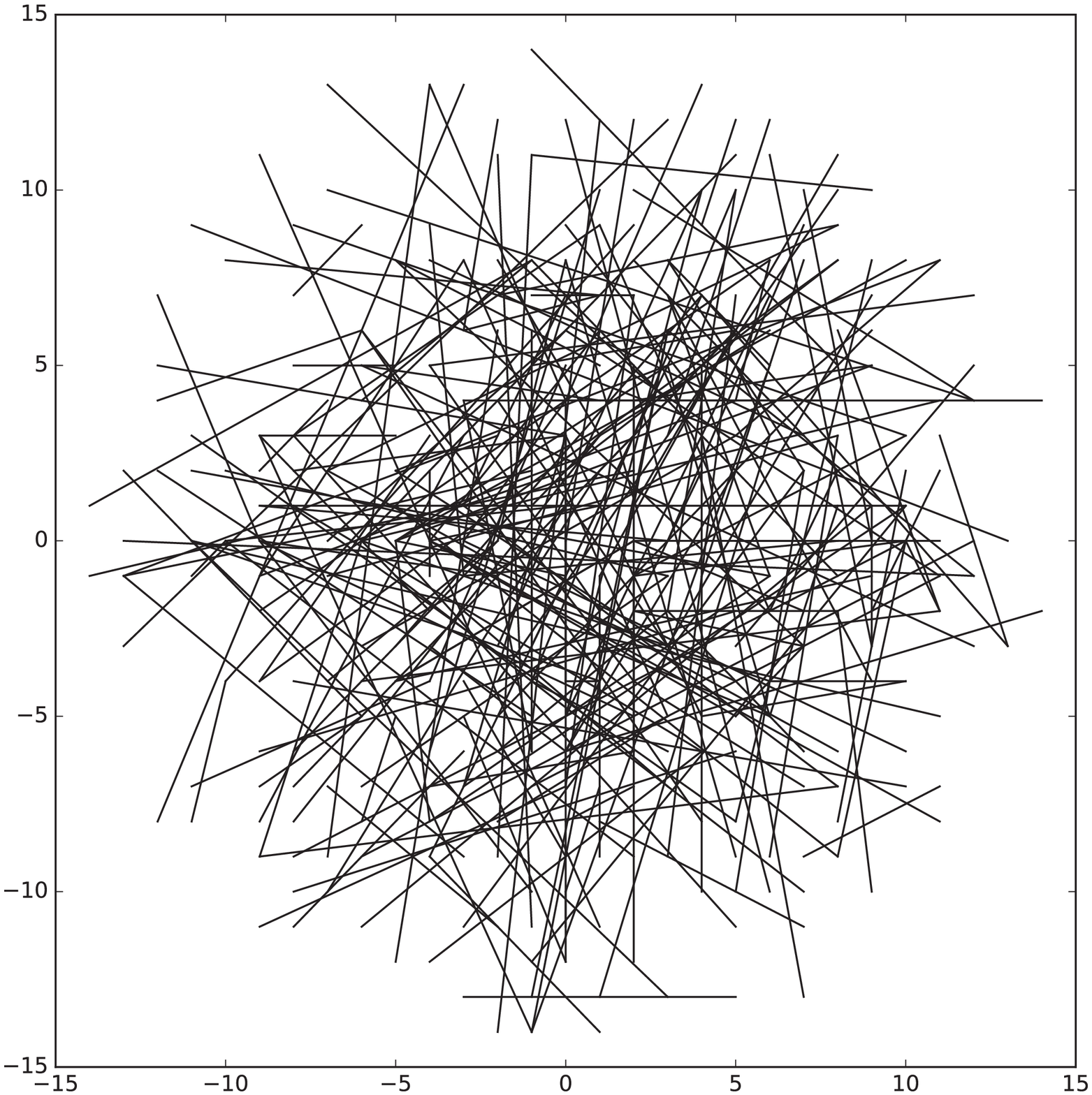}}
    \caption{Pattern of RS-BRIEF (left) and BRIEF (right).}
    \label{figure:fig2}
\end{figure}

Originally, $L_S$ and $L_D$ are randomly selected in the neighborhood according to Gaussian distribution. And every location after rotation needs to be calculated using the following formula:
\begin{equation}
\begin{aligned}
x' = x\cdot cos\theta - y \cdot sin \theta \\
y' = y\cdot cos\theta + x \cdot sin \theta
\end{aligned}
\end{equation}
where $(x,y)$ refers to the initial location and $(x',y')$ refers to the location after rotation. Since $512$ locations are required to be rotated in order to compute the descriptor of each feature, the rotation procedure is quite compute-intensive.

To reduce the computation cost of rotation procedure, a popular approach is to pre-compute the rotated BRIEF patterns \cite{Rublee2012ORB} instead of computing them directly each time. In this approach, the orientation of features is discretized into 30 different values, i.e., 12 degrees, 24 degrees, 36 degrees, etc. Then 30 BRIEF patterns after rotation are pre-computed and built as a lookup table. The lookup table is utilized to obtain the descriptors when necessary so that the computation cost could be reduced significantly.

One drawback of the above approach is the degradation in accuracy. Because the orientation of features is discretized, there will be a deviation from the true value which is up to 6 degrees (half of 12 degrees). However, considering that the test locations are selected from a circular patch with a radius of 15 pixels, the maximum error of a test location is about 1 pixel on the smoothened image. Hence, the influence on the accuracy is almost negligible.

Although the pre-computing approach could reduce the computation cost significantly in algorithm level, it is still difficult to implement them on hardware platforms directly. For FPGA hardware implementations, all the 30 BRIEF patterns are required to be pre-computed and stored as a lookup table, which will introduce considerable amount of extra resources so that it still could not satisfy the required energy efficiency.

In order to make descriptor computing more hardware-friendly, we put forward a special way to select the test locations and proposed a 32-fold rotationally symmetric BRIEF pattern (RS-BRIEF). The procedure to generate RS-BRIEF pattern is as follows. First of all, it selects 2 sets of locations, $L_{S1}(S_1,S_2, ...,S_8)$ and $L_{D1}(D_1,D_2, ...,D_8)$, in the neighborhood around the feature according to Gaussian distribution. Each of the 2 sets contains 8 locations. Then, it rotates $L_{S1}$ and $L_{D1}$ by increments of every 11.25 degrees, i.e., 11.25, 22.5, ..., 348.75, to generate  $L_{S2}, L_{S3},$ $..., L_{S32}$ and $L_{D2}, L_{D3}, ..., L_{D32}$. The 2 sets, $L_S'(L_{S1}\cup L_{S2}\cup ...\cup L_{S32})$ and $L_D'(L_{D1}\cup L_{D2}\cup ...\cup L_{D32})$, are the final test locations. The RS-BRIEF pattern is visualized and compared with the original BRIEF pattern in Figure \ref{figure:fig2}.

In summary, the rotationally symmetric pattern (RS-BRIEF) is generated by rotating the two sets of seeded locations, $L_{S1}$ and $L_{D1}$. To calculate descriptors with RS-BRIEF pattern, the operations of rotating test locations can be reduced to changing the order of these locations or shifting the generated descriptor. And consequently it could be much more hardware friendly than original BRIEF descriptors by dramatically reducing the computation without introducing extra memory footprint.


\section{\lowercase{e}SLAM ARCHITECTURE}\label{section:architecture}

\begin{figure}[htbp]
    \centering
    \includegraphics[width=0.3\textwidth]{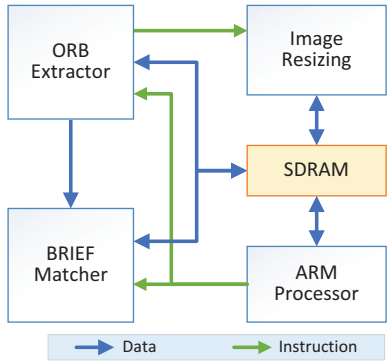}
    \caption{Overall architecture of \texttt{eSLAM}.}
    \label{figure:overall_architecture}
\end{figure}

The overall architecture of the proposed ORB-based visual SLAM accelerator, \texttt{eSLAM}, is shown in Figure \ref{figure:overall_architecture}. It is partially accelerated on programmable logic of FPGA and hosted by an ARM processor. The ORB Extractor and the BRIEF Matcher are implemented to accelerate feature extraction and matching, which account for over 90\% of the runtime on general computing platforms. And the Image Resizing module is adopted to generate image pyramids layer by layer for the ORB Extractor. When the ORB Extractor is processing one layer, the Image Resizing module applies nearest neighbor downsampling on the same layer to generate the next layer until the whole image pyramid is processed. The ARM processor performs pose estimation, pose optimization as well as map updating.

\subsection{ORB Extractor}

The ORB Extractor aims to extract ORB features from images. It reads data from SDRAM via AXI bus, and computes the ORB features with a local cache. After feature extraction is finished, it sends the result back to SDRAM and the descriptors of the features to the BRIEF Matcher. The original workflow of ORB feature extraction could be summarized as follows:

\begin{enumerate}[(1)]
\item \textbf{Detecting} keypoints from the input image. Assuming that $M$ keypoints are detected.
\item \textbf{Filtering} the keypoints. After filtering, only the $N$ keypoints with the best Harris scores are kept, where $N < M$.
\item \textbf{Computing} descriptors for the remained $N$ keypoints.
\end{enumerate}

Obviously there are two major problems when implementing the original workflow on hardware platforms. Firstly, the Detecting and Filtering procedures could be executed in parallel while the descriptors Computing procedure has to be idled until the Filtering is finished. Furthermore, it requires amount of on-chip cache to store the intermediate data when Computing the descriptors. In order to improve the computation throughput and reduce the memory consumption, the workflow of ORB feature extracting is rescheduled as a streaming manner as follows:

\begin{enumerate}[(1)]
\item \textbf{Detecting} keypoints from the input image. Assuming that $M$ keypoints are detected.
\item \textbf{Computing} descriptors for the detected $M$ keypoints.
\item \textbf{Filtering} and reserving $N$ features with the best Harris scores.
\end{enumerate}

After rescheduling, the descriptors Computing procedure is executed before Filtering procedure so that they could run simultaneously and be pipelined for the streaming keypoints. Compared with the original workflow, there are $M-N$ extra keypoints calculated which will introduce some overheads but the latency has been optimized significantly due to the eliminated idle states. Moreover, the required on-chip cache is also reduced dramatically according to the streaming processing manner.

\begin{figure}
    \centering
    \includegraphics[width=0.48\textwidth]{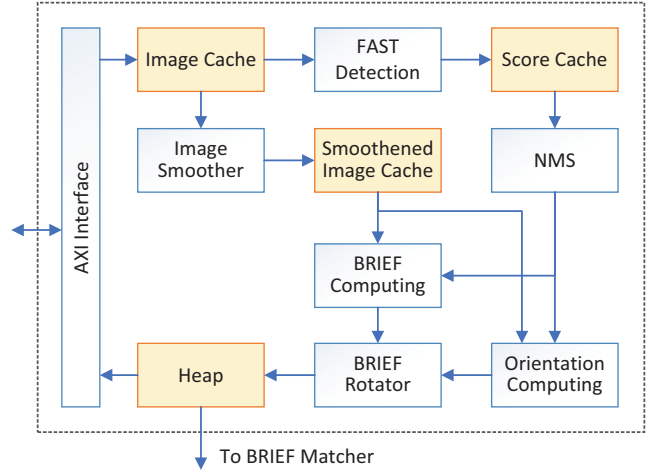}
    \caption{Architecture of the ORB Extractor.}
    \label{figure:orb_extractor}
\end{figure}

The detailed architecture of the ORB Extractor is shown in Figure \ref{figure:orb_extractor}. It is connected to AXI Interface and includes a FAST Detection module, a Image Smoother, an NMS (non-maximum suppression) module, a BRIEF Computing module, an Orientation Computing module, a BRIEF Rotator, a Heap and Caches (Image Cache, Score Cache and Smoothened Image Cache). The details of these modules are demonstrated as follows:

\textbf{AXI Interface:} The AXI Interface supports accessing SDRAM via AXI bus. The input image is read from SDRAM via AXI bus and stored in Image Cache while the computation results stored in the Heap are written back to SDRAM.

\textbf{FAST Detection:} The FAST Detection module takes a $7 \times 7$ pixels patch from the Image Cache as input. It detects FAST keypoint on this pixels patch and computes Harris corner score for each keypoint. If a FAST keypoint is detected, the corresponding Harris score is written into Score Cache.

\textbf{Image Smoother:} This module applies Gaussian blur operations on the $7\times 7$ pixels patch of the original image for smoothing. Then the smoothened image is utilized for calculating descriptors and orientations of features.

\textbf{NMS:} The NMS module applies non-maximum suppression on the results of the FAST Detection module. It removes FAST keypoints that are too close to each other, and only reserves the one with maximum Harris score in any $3\times 3$ pixels patch.

\textbf{Orientation Computing:} This module determines the orientation of each feature. The orientation is defined as the vector from the center of the feature to the mass center of the circular patch. The position $(u,v)$ of the mass center is defined as:
\begin{equation}
\begin{aligned}
u = \frac{\sum_{(x,y)\in C} \left( I(x,y)\cdot x \right)}{\sum_{(x,y)\in C}I(x,y)} \\
v = \frac{\sum_{(x,y)\in C} \left( I(x,y)\cdot y \right)}{\sum_{(x,y)\in C}I(x,y)}
\end{aligned}
\end{equation}
where $C$ refers to the circular patch and $I(x,y)$ refers to the intensity of the pixel located at $(x,y)$. The Orientation Computing module builds a lookup table to determine the orientation from $v/u$ and the signs of $u$ and $v$. Since the pattern of the test locations is 32-fold rotationally symmetric, the feature orientations are discretized and represented by an integral label ranged from 0 to 31, where 0 represents 0 degree, 1 represents 11.25 degrees, 2 represents 22.5 degrees, etc.

\textbf{BRIEF Computing:} The BRIEF Computing module takes circular patches of smoothened pixels to calculate descriptors for features. The test locations it uses to generate descriptors follow the rotationally symmetric pattern we proposed.

\textbf{BRIEF Rotator:} The BRIEF Rotator shifts the descriptor according to the feature orientation, which provides the same results as rotating the test locations of RS-BRIEF. Assuming that the feature orientation is $n$, the BRIEF Rotator moves the $8\times n$ bits from the beginning of the descriptor to the end.

\textbf{Heap:} The Heap is created to store and filter the descriptors, coordinates and Harris scores of features. To filter out some of the superfluous features, a max-heap structure is utilized to guarantee that only the 1024 features with the best Harris scores are reserved. Once the feature extraction is finished and stored in the heap, the descriptors and coordinates are sent to SDRAM through AXI Interface, and the descriptors are also delivered to the BRIEF Matcher.

\begin{figure}[htbp]
    \centering
    \includegraphics[width=0.48\textwidth]{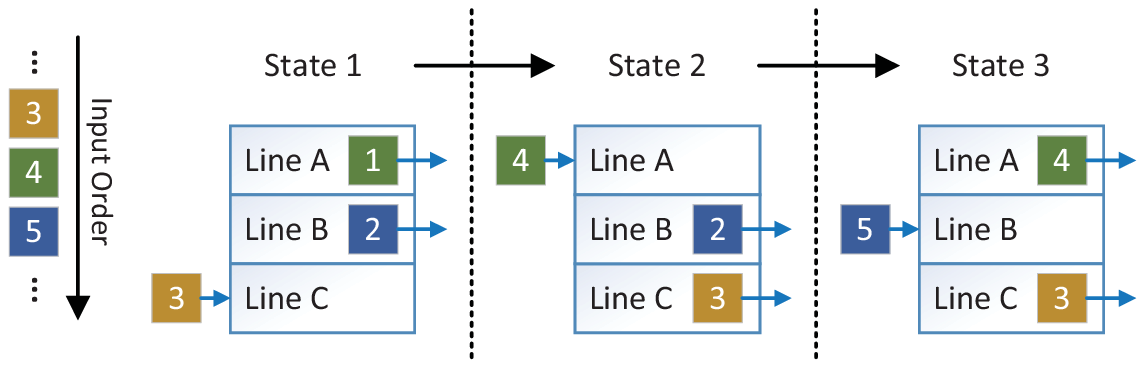}
    \caption{I/O mechanism of Image Cache. Line A, B and C refers to the 3 cache lines of Image Cache. Each square represents 8 columns of pixels.}
    \label{figure:cache_IO}
\end{figure}

\textbf{Cache:} There are 3 caches in ORB Extractor including the Image Cache storing pixels of the input image, the Score Cache storing the Harris scores of the keypoints, and the Smoothened Image Cache storing the smoothened image. These caches are designed by a manner of ``ping-pong mechanism'' so that the streaming data could be processed simultaneously. The Image Cache is taken as an example to explain the data I/O mechanism. The Image Cache consists of 3 cache lines, each of which stores 8 columns of image pixels. As shown in Figure \ref{figure:cache_IO}, the 3 cache lines receive input data by turns. The data I/O of the cache lines is controlled by a finite-state machine (FSM). The FSM is initialized by pre-storing 16 columns of pixels in cache line A and B. For each FSM state, one cache line receives input data while the other two send the data for output.

\subsection{BRIEF Matcher}

\begin{figure}[htbp]
    \centering
    \includegraphics[width=0.36\textwidth]{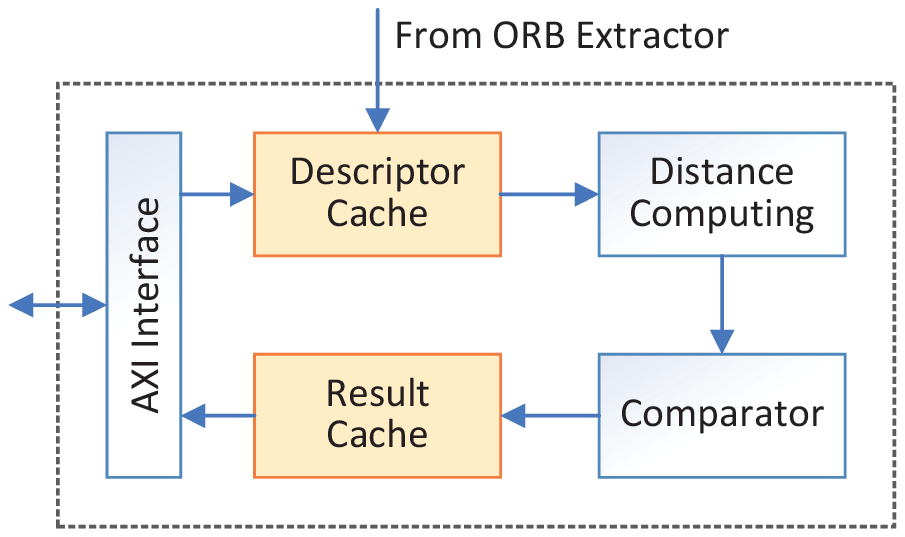}
    \caption{Architecture of the BRIEF Matcher.}
    \label{figure:brief_matcher}
\end{figure}

In the BREIF Matcher module, the features extracted from the current frame is compared with the map points of the global map. The features descriptors are obtained from the ORB Extractor, and the descriptors of global map are from SDRAM via AXI bus. The matching results are sent back to SDRAM at last.

The architecture of the BRIEF Matcher is shown in Figure \ref{figure:brief_matcher}. It is connected to AXI interface and includes a Descriptor Cache, a Distance Computing module, a Comparator and a Result Cache. The matching procedure starts following the ORB extraction. Assuming 2 sets of descriptors $D_A({DS}_1,{DS}_2,$ $...,{DS}_n )$ and $D_B({DD}_1,{DD}_2,$ $...,{DD}_m)$ have been pre-stored in Descriptor Cache, where $D_A$ is the descriptors obtained from current frame, and $D_B$ is the descriptors of the map points in the global map. For each descriptor ${DS}_i$ in $D_A$, the Distance Computing module calculates the Hamming distances between ${DS}_i$ and each descriptor ${DD}_j$ in $D_B$. With the calculated Hamming distances ${HD}(H_{i1},H_{i2},$ $...,H_{im} )$, the Comparator searches through ${HD}$ and finds the minimum value to determine the matching result and stores them into the Result Cache.

\subsection{Parallelizing Mechanism}

\begin{figure}
    \centering
    \includegraphics[width=0.48\textwidth]{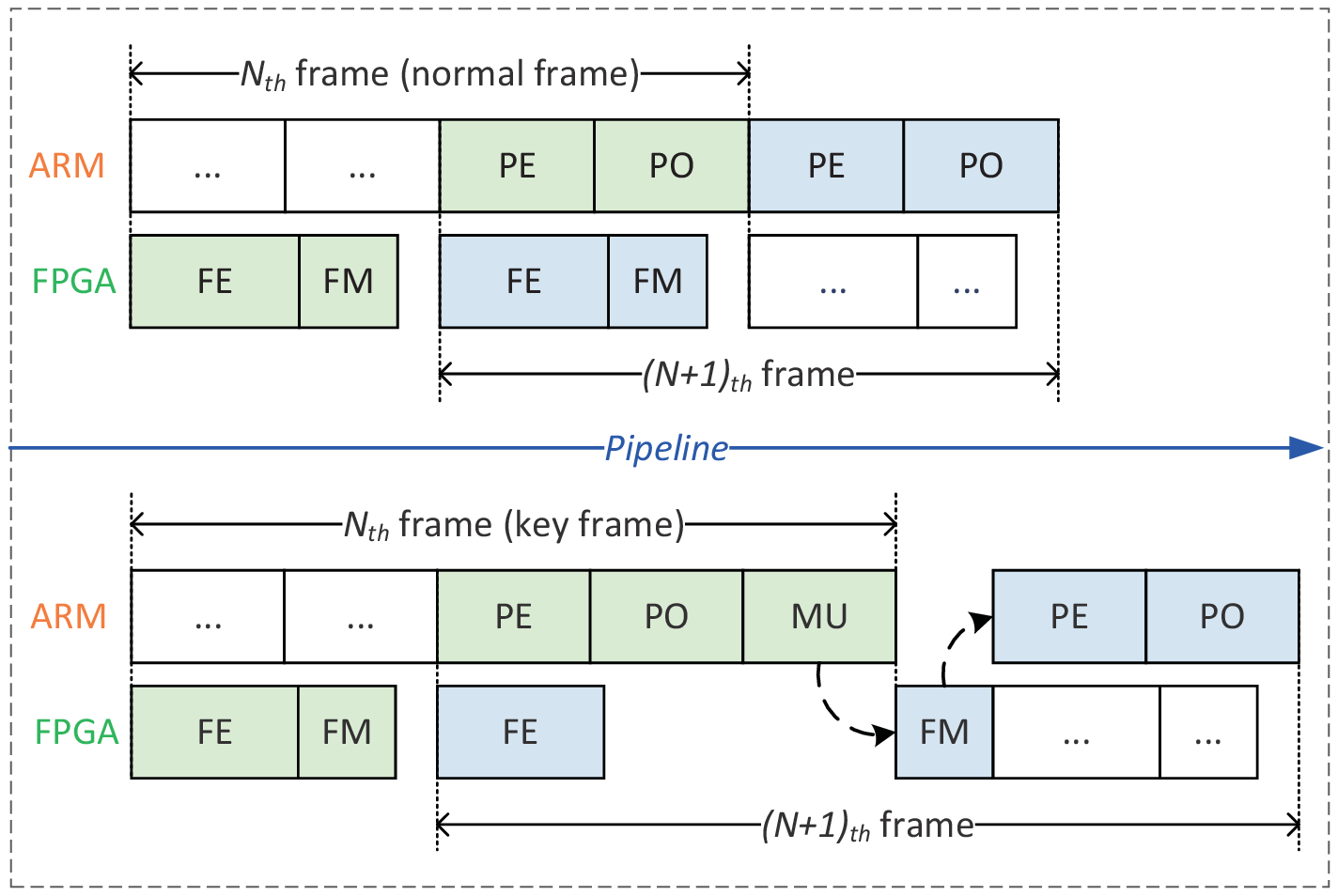}
    \caption{Parallelized pipeline of normal frame (upper) and key frame (lower), where FE refers to feature extraction, FM refers to feature matching, PE refers to pose estimation, PO refers to pose optimization and MU refers to map updating.}
    \label{figure:pipeline}
\end{figure}

Since \texttt{eSLAM} is a heterogeneous system with the ARM processor as the host controller and FPGA as the acceleration modules, the parallelizing mechanism is critical to improve the computation throughput. The utilized parallelized pipeline is shown in Figure \ref{figure:pipeline}. For normal frames processing, while the ARM processor is performing pose estimation and pose optimization, the ORB Extractor and BRIEF Matcher are fired up to do feature extraction and feature matching for the next frame.
However, it is different to process key frames because map updating is executed on the ARM processor after pose estimation and pose optimization. The ORB Extractor performs feature extraction on FPGA in parallel with the ARM processor, but the BRIEF Matcher would not start to work until map updating is finished.

With the parallelizing mechanism above, the several stages could be performed efficiently in pipeline. For normal frames, feature extraction and matching runs in parallel with pose estimation and optimization. And for key frames, feature extraction runs in parallel with pose estimation and optimization. These parallel processing manners could improve the computing throughout significantly.

\section{EXPERIMENTAL RESULTS}\label{section:evaluation_and_results}

\subsection{Experimental Setup}

\textbf{Hardware Implementation:} The proposed \texttt{eSLAM} system is implemented on Xilinx Zynq XCZ7045 SoC~\cite{zynq}, which integrates an ARM Cortex-A9 processor and FPGA resources. The clock frequency of the ARM processor is 767 $MHz$, and the clock of accelerating modules is 100 $MHz$. The resource utilization of the proposed system is shown in Table \ref{tab:tab1}. Since only about 1/4 resources are utilized on XCZ7045, it is possible to prototype them onto SoCs with less resources and lower price, such as XCZ7030/XCZ7020.

\begin{table}[htb]
\centering
\caption{The FPGA resources utilization of \texttt{eSLAM}.}
\label{tab:tab1}
\begin{tabular}{l|c|c|c|c}
\hline
  & LUT     & FF   & DSP     & BRAM          \bigstrut                \\ \hline
\multicolumn{1}{c|}{Utilization} & \begin{tabular}[c]{@{}c@{}}56954   \\
(26.0\%)\end{tabular}            & \begin{tabular}[c]{@{}c@{}}67809  \\
(15.5\%)\end{tabular}            & \begin{tabular}[c]{@{}c@{}}111    \\
(12.3\%)\end{tabular}            & \begin{tabular}[c]{@{}c@{}}78     \\
(14.3\%)\end{tabular} \bigstrut \\ \hline
\end{tabular}
\end{table}

\textbf{Dataset:} The proposed \texttt{eSLAM} is evaluated on TUM dataset \cite{sturm2012benchmark}. It contains RGB images along with depth information and is widely used in visual SLAM community. The image resolution is $640\times480$. Five different sequences in the dataset, $fr1/xyz$, $fr1/desk$, $fr1/room$, $fr2/$ $xyz$ and $fr2/rpy$ are used for evaluation. Each sequence contains a ground truth trajectory that is obtained by a high-accuracy motion-capture system.

\subsection{Accuracy Analysis}

\begin{figure}
    \centering
    \includegraphics[width=0.42\textwidth]{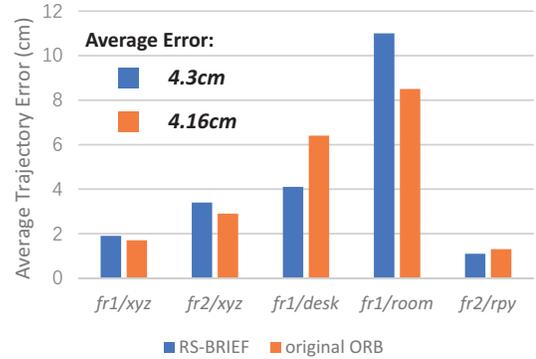}
    \caption{Average trajectory error of the SLAM implementation with RS-BRIEF compared with the original ORB on TUM dataset.}
    \label{figure:ate}
\end{figure}

The accuracy of the visual SLAM system is measured by trajectory error which means the difference between the ground truth trajectory and the estimated trajectory. As shown in Figure \ref{figure:ate}, the average trajectory error is compared with the original ORB based SLAM implementation on the five sequences from TUM dataset. For $fr1/xyz$, $fr1/room$, and $fr2/xyz$ sequence, the implementation with original ORB has a better accuracy than with RS-BRIEF descriptor. However, the implementation with RS-BRIEF descriptor could have a better accuracy than with original ORB when evaluated on $fr1/desk$ and $fr2/rpy$ sequence. Among the five sequences, the total average error of RS-BRIEF based implementation is about $4.3$ $cm$, and the original ORB based implementation is about $4.16$ $cm$, which indicates that the accuracy of RS-BRIEF descriptor is comparable to the original descriptor.

\begin{figure}
    \centering
    \includegraphics[width=0.4\textwidth]{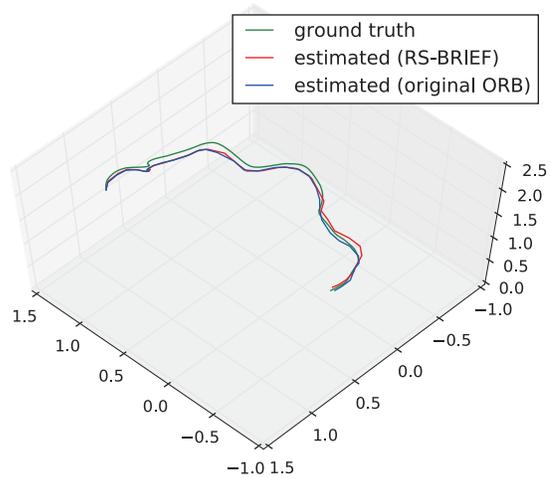}
    \caption{Estimated trajectory of the RS-BRIEF descriptor and original ORB descriptor based SLAM implementations, compared with the ground truth trajectory on $fr1/desk$.}
    \label{figure:trajectory}
\end{figure}

Meanwhile, the trajectories estimated by the RS-BRIEF based implementation and the original ORB based implementation are also compared with the ground truth trajectory on $fr1/desk$ sequence and visualized in Figure \ref{figure:trajectory}. Aiming to display the trajectories clearly, only a piece of them are selected as shown in Figure \ref{figure:trajectory}.

\subsection{Performance Evaluation}


The performance of the proposed \texttt{eSLAM} system is compared with the software implementations on the integrated ARM Cortex-A9 processor of XCZ7045 SoC and an Intel i7-4700mq processor~\cite{intelcpu}. The runtime comparison is shown in Table \ref{tab:tab2}. Accelerated by ORB Extractor and BRIEF Matcher, the latency of feature extraction and matching procedure in \texttt{eSLAM} is reduced to $9.1$ $ms$ and $4$ $ms$, respectively. Compared with Intel CPU and ARM, \texttt{eSLAM} could achieve $3.6\times$ and $32\times $ speedup in feature extraction, $4.9\times $ and $61.6\times $ speedup in feature matching.

\begin{table}
\centering
\caption{Detailed runtime breakdown of \texttt{eSLAM} compared with software implementations on ARM processor and Intel i7 CPU.}
\label{tab:tab2}
\begin{tabular}{l|c|r|r}
\hline
\multicolumn{1}{c|}{} & \texttt{eSLAM}     & \multicolumn{1}{c|}{ARM} & \multicolumn{1}{c}{Intel i7}                                                  \bigstrut \\ \hline
Feature Extraction    & \multicolumn{1}{r|}{9.1 $ms$}  & 291.6 $ms$ & 32.5 $ms$                                                   \bigstrut \\ \hline
Feature Matching      & \multicolumn{1}{r|}{4.0 $ms$}  & 246.2 $ms$ & 19.7 $ms$                                                   \bigstrut \\ \hline
Pose Estimation       & \multicolumn{2}{c|}{9.2 $ms$}  & 0.9 $ms$                                                    \bigstrut \\ \hline
Pose Optimization     & \multicolumn{2}{c|}{8.7 $ms$}  & 0.5 $ms$                                                    \bigstrut \\ \hline
Map Updating          & \multicolumn{2}{c|}{9.9 $ms$}  & 1.2 $ms$                                                    \bigstrut \\ \hline
\end{tabular}
\end{table}


Table \ref{tab:tab3} compares the average runtime per frame, the frame rate, the energy consumed per frame, the power consumption of \texttt{eSLAM} with the ARM processor and the Intel CPU. For normal frames, \texttt{eSLAM} performs feature extraction (FE) and matching (FM) simultaneously with pose estimation (PE) and optimization (PO). The average runtime is the sum of processing time of PE and PO, $17.9$ $ms$. For key frames, FE is performed simultaneously with PE. \texttt{eSLAM}'s average runtime time is $31.8$ $ms$, which is the sum of processing time of FM, PE, PO and MU. Compared with the ARM processor, \texttt{eSLAM} achieves about $17.8\times$ speedup when processing key frames and $31\times$ speedup for normal frames. Compared with the Intel i7 processor, it could achieve $1.7\times$ to $3\times$ speedups.

In terms of energy consumption, the proposed \texttt{eSLAM} also shows great advantage compared with the ARM and Intel CPU. Although the power consumption of \texttt{eSLAM} is increased by about $23\%$ compared with the ARM processor due to the additional FPGA accelerating modules, the energy consumed per frame is still reduced by $14\times$ to $25\times$ depending on the key frame rate. Compared with the Intel i7 processor, the energy consumption is reduced by $41\times$ to $71\times$.

\subsection{Discussions}

As shown in Table \ref{tab:tab3}, the key frame rate of \texttt{eSLAM} is $31.45$ $fps$, and the normal frame rate is $55.87$ $fps$, which is much less than $171$ $fps$ which is achieved by Navion~\cite{suleiman2018navion}. This gap is mainly because of the adopted different algorithms. Navion adopts the optical-flow method while only keypoints are detected but descriptors calculation and feature matching are not required. However, the adopted feature-based approach in \texttt{eSLAM} is much more robust in many scenarios where optical-flow methods may fail. Because the optical-flow methods are only available with two basic assumptions: constant illumination and small motions/displacements existed \cite{fleet2006optical}.

\begin{table}
\centering
\caption{Frame rate and energy efficiency comparison results, where ``N-frame'' represents the normal frame, and ``K-frame'' represents the key frame.}
\label{tab:tab3}
\begin{tabular}{cc|r|r|r}
\hline
                                                                                                &         & \multicolumn{1}{c|}{ARM}      & \multicolumn{1}{c|}{Intel i7}  & \multicolumn{1}{c}{\texttt{eSLAM}}   \bigstrut \\ \hline
\multicolumn{1}{c|}{\multirow{2}{*}{Runtime}} & N-frame & 555.7 $ms$ & 53.6 $ms$   & 17.9 $ms$                                                     \bigstrut \\ \cline{2-5}
\multicolumn{1}{c|}{}                                                                           & K-frame & 565.6 $ms$ & 54.8 $ms$   & 31.8 $ms$                                                    \bigstrut \\ \hline
\multicolumn{1}{c|}{\multirow{2}{*}{\begin{tabular}[c]{@{}c@{}}Frame\\ Rate\end{tabular}}}      & N-frame & 1.8 $fps$  & 18.66 $fps$ & 55.87 $fps$                                                  \bigstrut \\ \cline{2-5}
\multicolumn{1}{c|}{}                                                                           & K-frame & 1.77 $fps$ & 18.25 $fps$ & 31.45 $fps$                                                  \bigstrut \\ \hline
\multicolumn{2}{c|}{Power}                                                                                & 1.574 $W$  & 47 $W$      & 1.936 $W$                                                    \bigstrut \\ \hline
\multicolumn{1}{c|}{\multirow{2}{*}{\begin{tabular}[c]{@{}c@{}}Energy\\ per Frame\end{tabular}}}                                                    & N-frame & 875 $mJ$   & 2519 $mJ$   & 35 $mJ$                                                      \bigstrut \\ \cline{2-5}
\multicolumn{1}{c|}{}                                                                           & K-frame & 890 $mJ$   & 2575 $mJ$   & 62 $mJ$                                                      \bigstrut \\ \hline
\end{tabular}
\end{table}


Compared with the ORB extractor implemented on FPGA in \cite{Fang2018FPGA}, the ORB extractor in \texttt{eSLAM} has deployed hardware-friendly optimization, such as RS-BRIEF and workflow rescheduling. Hence, the latency of feature extraction in \texttt{eSLAM} is approximately $39\%$ less than the latency of \cite{Fang2018FPGA}, even if $48\%$ more pixels are processed in \texttt{eSLAM} because of the involved extra two layers in the image pyramid.

\section{CONCLUSIONS}\label{section:conclusion}

In this paper, a heterogeneous ORB-based visual SLAM system, \texttt{eSLAM}, is proposed for energy-efficient and real-time applications and evaluated on Zynq platforms. The ORB algorithm is first reformulated as a rotationally symmetric pattern for hardware-friendly implementation. Meanwhile, the most time-consuming stages, i.e., feature extraction and matching, are accelerated on FPGA to reduce the latency significantly. The \texttt{eSLAM} is also designed as a pipelined manner to further improve the throughput and reduce the memory footprint. The evaluation results on TUM dataset have shown \texttt{eSLAM} could achieve $1.7\times $ to $3\times $ speedup in frame rate, and $41\times$ to $71\times$ improvement in energy efficiency when compared with the Intel i7 CPU. Compared with the ARM processor, \texttt{eSLAM} could achieve $17.8\times$ to $31\times$ speedup in frame rate, and $14\times$ to $25\times$ improvement in energy efficiency.



\end{document}